\documentclass[12pt]{article}
\usepackage{epsfig}
\usepackage[centerlast,footnotesize]{caption}
\begin{document}

\begin{center}
{\Large \bf Electromagnetic properties of non-Dirac particles with rest spin
1/2}

\vspace{0.5 cm}

\begin{small}
\renewcommand{\thefootnote}{*}
L.M.Slad\footnote{slad@theory.sinp.msu.ru} \\

\vspace{0.3 cm}

{\it Skobeltsyn Institute of Nuclear Physics,
Lomonosov Moscow State University, Moscow 119991, Russian Federation}
\end{small}
\end{center}

\vspace{0.5 cm}

\begin{footnotesize}
We resolve a number of questions related to an analytic description of electromagnetic form factors of non-Dirac particles with the rest spin 1/2. We find the general structure of a matrix antisymmetric tensor operator. We obtain two recurrence relations for matrix elements of finite transformations of the proper Lorentz group and explicit formulas for a certain set of such elements. Within the theory of fields with double symmetry, we discuss writing the components of wave vectors of particles in the form of infinite continued fractions. We show that for $Q^{2} \leq 0.5$ (GeV/c)$^{2}$ , where $Q^{2}$ is the transferred momentum squared, electromagnetic form factors that decrease as  $Q^{2}$ increases and are close to those experimentally observed in the proton can be obtained without explicitly introducing an internal particle structure.
\end{footnotesize}

\vspace{0.5 cm}

\begin{small}

\begin{center}
{\large \bf 1. Introduction}
\end{center}

The difference between the proton and the electron that shows up in their electromagnetic interaction was first evidenced in the 1933 measurement of the magnetic moment of the proton [1] and was finally established after the McAllister and Hofstadter experiment on elastic scattering of electrons on protons [2]. The essence of this difference is that one or more of the following three electron characteristics are inapplicable to the proton. First, the electron is assigned a Dirac representation of the proper Lorentz group 
$L^{\uparrow}_{+}$. Second, up to the experimentally attainable sizes of the order $10^{-16}$ cm, the electron behaves as a pointlike particle, which corresponds to the locality of the Lagrangian describing its interaction with the photon, where the fields of all particles are taken at the same point. Third, the Lagrangian of the electromagnetic interaction of the electron is minimal: expressed in terms of the vector potential $A^{\mu}$ of the electromagnetic field, it is free of derivatives.
      The theoretical concept of the proton and neutron that subsequently formed appeals to their nonminimal electromagnetic interaction and to a complicated internal structure but hardly challenges the assumption that the nucleon is a Dirac particle. This is reflected in the general form of the nucleon electromagnetic current in the momentum space [3]:
\begin{equation}
j^{\mu}(p,p_{0}) = ie \bar{u}(p) \left[ \gamma^{\mu} F_{1}(Q^{2}) +
\frac{i\kappa}{2M} \sigma^{\mu\nu} q_{\nu} F_{2}(Q^{2}) \right] u(p_{0}),
\end{equation}
where $\bar{u}$ and $u$ are Dirac spinors, $\kappa$ is the anomalous magnetic moment, $M$ is the nucleon mass, $q = p - p_{0}$, and $Q^{2} = -q^{2}$. Along with the Dirac ($F_{1}$) and Pauli ($F_{2}$) form factors in (1), the Sachs form factors $G_{E}$ and $G_{M}$ are introduced as [4]
\begin{equation}
G_{E} = F_{1} - \kappa \tau F_{2}, \qquad G_{M} = F_{1} + \kappa F_{2},
\end{equation}
where $\tau = Q^{2}/4 M^{2}$; these form factors are believed to describe the distribution of the electric charge and the magnetism of the nucleon. Using the quantities $G_{E}$ and $G_{M}$ brings the Rosenbluth formula [5] for the elastic scattering of nonpolarized electrons on nonpolarized nucleons in the laboratory frame to the simplest form
\begin{equation}
\frac{d\sigma}{d\Omega} = \frac{\alpha^{2} E' \cos^{2}(\theta/2)}
{4 E^{3}\sin^{4}(\theta/2)} \left[
\frac{G_{E}^{2}+\tau G_{M}^{2}}{1+\tau} + 2\tau G_{M}^{2} 
\tan^{2}\frac{\theta}{2} \right] ,
\end{equation}
where $E$ and $E'$ are the respective electron energies in the initial and final states and $\theta$ is the electron scattering angle. A detailed review of the results of experimental measurements of electromagnetic form factors of the proton and the neutron and their various theoretical interpretations can be found in [6].

Rejecting the description of the proton by a Dirac spinor entails numerous indeterminancies. One type of indeterminancy arises because a particle with the rest spin 1/2 can be assigned an infinite number of irreducible representations of the proper Lorentz group $L^{\uparrow}_{+}$ of the 
form\renewcommand{\thefootnote}{1)}\footnote{Irreducible representations of the L group are labeled [7], [8] by two indices $(l_{0}, l_{1})$, where $2l_{0}$ is an integer and $l_{1}$ is an arbitrary complex number. The canonical basis of the $(l_{0}, l_{1})$-representation space is related to the $SO(3)$ subgroup
and is denoted by $\xi_{(l_{0}, l_{1})lm}$, where $l$ is the spin and $m$ is its projection on the third axis, with $m = -l, -l+1, \ldots, l$ and
$l = |l_{0}|, |l_{0}|+1, \ldots$. In general, the range of values of $l$ is infinite. The $(l_{0}, l_{1})$ representation is finite dimensional, and the
above sequence of spins terminates at the number $|l_{1}|-1$ if $2l_{0}$ and 
$2l_{1}$ are integers of the same parity and $|l_{1}| > |l_{0}|$. The Dirac representation is $(-1/2, 3/2) \oplus (1/2, 3/2)$.} $(-1/2, l_{1})$ and 
$(1/2, l_{1})$, where $l_{1}$ is an arbitrary complex number, and infinitely many reducible representations constructed from these. Another type of indeterminancy arises from the arbitrariness in the field theory constants allowed by the relativistic invariance [7], [8], and this arbitrariness can be infinite, for example, if ISFIR-class fields are considered, which transform under representations decomposable into an infinite direct sum of 
finite-dimensional irreducible representations of the $L^{\uparrow}_{+}$ group.
                                                   
Arguments in favor of the possible description of hadrons by ISFIR-class fields were adduced in [9], [10]. We note that these fields had not been investigated until the appearance of a symmetry approach for eliminating the infinite arbitrariness in the corresponding Lagrangians. The general definition of the double symmetry notion [11], including the Gell-Mannб╜--Levy $\sigma$-model symmetry [12] and supersymmetry as particular cases, allowed supplementing [9] the relativistic invariance (the primary symmetry) of ISFIR-class field theories with the requirement of their additional (secondary) symmetry generated by transformations of the form
\begin{equation}
\psi (x) \rightarrow \psi '(x) = \exp [-i D^{\mu} \theta_{\mu}] \psi (x),
\end{equation}
where the $D^{\mu}$ are matrix operators and the parameters  $\theta_{\mu}$ are components of a polar or an axial four-vector of the orthochronous Lorentz group $L^{\uparrow}$. To avoid infinite degeneracy with respect to spin in the mass
spectrum of the resulting theory because the Lorentz group is extended, spontaneous secondary symmetry breaking is introduced [13]. As a result, as shown in [14], the free relativistic invariant equations for ISFIR-class fermionic fields thus constructed yield mass spectra that agree with the experimental picture of baryon resonances and with the parton model of hadrons supplemented by the confinement hypothesis. 

We note that all the previously considered relativistically invariant free field theories with infinitely many degrees of freedom, namely, bilocal equations (see [15]) and Gelfandб╜--Yaglom-type equations [7], [8] (with the corollaries refined in [16]) for FSIIR-class fields transforming under representations decomposable
into a finite direct sum of infinite-dimensional irreducible representations of the $L^{\uparrow}_{+}$ group, have mass spectra that are unsuitable for particle physics because of an accumulation point at zero. The infinite number of states in the theory of ISFIR-class fields with double symmetry seems to reflect some internal structure of the corresponding particles, and the physically entirely satisfactory mass spectra of the theory are evidence that the proposed approach describing a hadron by an infinite-component monolocal field is essentially correct. It is now desirable to find out how the physical content of the existing structure characteristics of particles with rest spin 1/2 is affected by passing from the Dirac representation of the $L^{\uparrow}_{+}$ group to a non-Dirac representation assigned to the particle as a whole and whether a satisfactory description of experimental processes involving composite particles can be obtained based on considering a local interaction of monolocal fields without explicitly introducing structural quantities.

\begin{center}
{\large \bf 2. General description of the electromagnetic current \\ 
and form factors of non-Dirac particles with rest spin 1/2}
\end{center}

The first general results concerning the electromagnetic properties of a nucleon that follow from rejecting its description with a Dirac spinor were recently obtained in [10]. For a broad class of representations of the proper Lorentz group assigned to a field $\psi$ of a particle with rest spin 1/2 and mass $M$, a polar four-vector of the electromagnetic current free of structure functions of the transferred momentum squared can be written as
\begin{equation}
{\cal J}^{\mu}(p, p_{0}) = ie \left( \psi(p), \; \left[ Q_{0} \Gamma^{\mu} + 
M^{-1}\Lambda^{\mu\nu}(p,p_{0}) q_{\nu} \right] \psi(p_{0}) \right) ,
\end{equation}
where
\begin{equation}
\Lambda^{\mu\nu}(p,p_{0}) \equiv \Gamma^{\mu\nu}
+M^{-1}a_{i_{1}}\Gamma^{\mu\nu\nu_{1}}(p_{i_{1}})_{\nu_{1}}+ \ldots 
+M^{-j}a_{i_{1}\ldots i_{j}}\Gamma^{\mu\nu\nu_{1}\ldots \nu_{j}}(p_{i_{1}})_{\nu_{1}} \ldots (p_{i_{j}})_{\nu_{j}} + \ldots ,
\end{equation}
$(\psi_{1}, \psi_{2})$ is a relativistically invariant bilinear form assumed to be nondegenerate, $Q_{0}$ is the particle charge in units of the positron charge $e$, $\Gamma^{\mu}$ is the four-vector matrix operator, $\Gamma^{\mu\nu}$ and 
$\Gamma^{\mu\nu\nu_{1}\ldots \nu_{j}}$ are traceless matrix tensor operators of the Lorentz group that are antisymmetric in $\mu$ and $\nu$, 
$a_{i_{1}\ldots i_{j}}$ are numerical coefficients, and 
$p_{k} \in \{ p_{0}, p \}$ for any index $k$. We assume that the field vector 
$\psi(p)$ satisfies some relativistically invariant equation
\begin{equation}
(\Gamma^{\mu} p_{\mu} - R)\psi(p) = 0,
\end{equation}
where $R$ is a scalar matrix operator.

The terms in relations (5) and (6) involving the four-momentum $p_{k}$ correspond to terms with derivatives of the field $\psi (x)$ in the coordinate representation of the electromagnetic interaction Lagrangian and can
then be treated as higher electric multipoles of a composite particle. It is more convenient to consider the set $\{ p_{0}, q \}$ instead of the set of four-momenta $\{ p_{0}, p \}$ in (5) and (6).

It was shown in [10] that the angular distribution of final particles in the process of elastic scattering of nonpolarized electrons on a nonpolarized particle with rest spin 1/2, independently of the $L^{\uparrow}_{+}$-group 
representation $S_{0}$ assigned to them, is given by Rosenbluth formula (3), where the role of the electromagnetic factor is played by the quantities
\begin{equation}
G_{E} = \frac{C}{\sqrt{\tau +1}} 
(\psi_{+1/2}(p), \; [Q_{0}R-q^{3}\Lambda^{03}] \psi_{+1/2}(p_{0})),
\end{equation}
\begin{equation}
G_{M} = \frac{C}{\sqrt{\tau}} 
(\psi_{+1/2}(p), \; [Q_{0}M\Gamma^{1}+q^{0}\Lambda^{10}-
q^{3}\Lambda^{13}] \psi_{-1/2}(p_{0})) ,
\end{equation}
where
\begin{equation}
C = (\psi_{+1/2}(p_{0}), \; R \psi_{+1/2}(p_{0}))^{-1}.
\end{equation}

The quantities $G_{E}$ and $G_{M}$ given by these formulas are constants for Dirac particles and nontrivial functions of the transferred momentum squared 
$Q^{2}$ for non-Dirac particles with rest spin 1/2. The specific form of the functions $G_{E}(Q^{2})$ and $G_{M}(Q^{2})$ is determined by several circumstances: first, by the choice of the representation $S_{0}$; second, by the decomposition of the wave vector $\psi(p_{0})$ of the particle in its rest frame with respect to the canonical basis vectors in the representation space 
$S_{0}$, which depends on the structure of the operators in Eq. (7); third, by the constants of the operators $\Gamma^{\mu\nu}$ and 
$\Gamma^{\mu\nu\nu_{1}\ldots \nu_{j}}$, $j=1,2,\ldots$. None of these circumstances depends explicitly on the internal structure of a non-Dirac particle with rest spin 1/2, and we can therefore say that the corresponding form factors $G_{E}$ and $G_{M}$ in Rosenbluth formula (3) are given by the external characteristics of the particle as a whole.

We next provide a description, which is missing from the literature, of several quantities in the general relativistic field theory that are needed for calculating the form factors of non-Dirac particles with rest spin 1/2 in (8) and (9). We find the general structure of second-rank matrix antisymmetric tensor operators and explicitly give some matrix elements of finite transformations of the proper Lorentz group for finite-dimensional irreducible representations $(\pm 1/2, l_{1})$; such elements were previously found only for unitary (infinite-dimensional) irreducible representations [17]. Along with this, we consider problems related to the current in (5) and (6), whose solutions are given in the framework of the theory of infinite-component
ISFIR-class fields with double symmetry. Specifically, we give the results connected with eliminating the infinite arbitrariness in the constants of the  
$\Gamma^{\mu\nu}$ operator, discuss the structure of tensor operators 
$\Gamma^{\mu\nu\nu_{1}\ldots \nu_{j}}$, $j=1,2,\ldots$ of rank three and higher, and propose writing the components of wave vectors satisfying Eq. (7) in the form of infinite continued fractions.

In the theory of ISFIR-class fields with double symmetry, electromagnetic current (5), (6) contains an infinite number of matrix operators in the general case and hence an infinite number of arbitrary (normalization) constants. Such a description of the electromagnetic interaction of non-Dirac particles lacks
simplicity and elegance. So far, we can see no symmetry-based way to eliminate the infinite arbitrariness in the constants of this current. Nonetheless, restricting ourself in current (5), (6) to tensor operators of rank four and to values of the transferred momentum squared not exceeding 0.5 (GeV/c)$^{2}$, we use numerical computations to conclude that we can in principle obtain electromagnetic form factors that decrease as $Q^{2}$ increases, without explicitly involving an internal structure of particles with rest spin 1/2.

\begin{center}
{\large \bf 3. General structure of the second-rank  \\ matrix antisymmetric
tensor operators}
\end{center}

Let the transformation
\begin{equation}
x'_{\mu} = {[l(g)]_{\mu}}^{\nu} x_{\nu}. 
\end{equation}
of covariant four-vector components correspond to an element $g$ of the proper Lorentz group. Then the corresponding transformation of covariant components of a second-rank antisymmetric tensor can be written as
\begin{equation}
y'_{\mu\nu} ={[T(g)]_{\mu\nu}}^{\alpha\beta} y_{\alpha\beta},
\end{equation}
where
\begin{equation}
{[T(g)]_{\mu\nu}}^{\alpha\beta} = \frac{1}{2} \{ {[l(g)]_{\mu}}^{\alpha}
{[l(g)]_{\nu}}^{\beta} - {[l(g)]_{\nu}}^{\alpha} {[l(g)]_{\mu}}^{\beta} \} .
\end{equation}

Matrix operators $K^{\mu\nu}$ that act on the field $\psi$ as on a vector in a representation space $S(g)$ of the proper Lorentz group constitute an antisymmetric operator if the quantity $(\psi_{1}, K^{\mu\nu} \psi_{2})$ transforms under the $L^{\uparrow}_{+}$ group as an antisymmetric tensor, i.e., the contraction $(\psi_{1}, K^{\mu\nu} \eta_{\mu\nu}\psi_{2})$ is an invariant nondegenerate bilinear form. This implies that the operators $K^{\mu\nu}$ must satisfy the condition
\begin{equation}
K^{\mu\nu} = S^{-1}(g) K^{\alpha\beta} S(g) {[T(g)]_{\alpha\beta}}^{\mu\nu}.
\end{equation}

An infinitesimal proper Lorentz transformation of a four-vector and a vector in the $S(g)$-representation space are respectively introduced as [7]
\begin{equation}
x'_{\mu} = x_{\mu} + {\epsilon_{\mu}}^{\nu}x_{\nu},
\qquad
\psi' = \psi + \frac{1}{2} \epsilon_{\mu\nu} I^{\mu\nu} \psi,
\end{equation}
where the parameters of the transformations $\epsilon_{\mu\nu}$ and the infinitesimal operators $I^{\mu\nu}$ are antisymmetric: 
$\epsilon_{\mu\nu} = -\epsilon_{\nu\mu}$ and $I^{\mu\nu} = -I^{\nu\mu}$.
               
It follows from formulas (11) and (13)--(15) that the commutation relations
\begin{equation}
[I^{\mu\nu}, K^{\alpha\beta}] = -g^{\mu\alpha}K^{\nu\beta}+ 
g^{\mu\beta}K^{\nu\alpha} + g^{\nu\alpha}K^{\mu\beta} - 
g^{\nu\beta}K^{\mu\alpha},
\end{equation}
where $g^{00}=-g^{11}=-g^{22}=-g^{33}=1$ and $g^{\gamma\delta}=0$ for 
$\gamma \neq \delta$.
              
We note that the commutation relations for infinitesimal operators follow from (16) under the replacement $K^{\gamma\delta} \rightarrow I^{\gamma\delta}$. The matrix realization of the operators $I^{\mu\nu}$ in the space of any irreducible representation of the $L^{\uparrow}_{+}$ group is known (see [7], [8], or [9]).
        
Among the 36 conditions in (16), only eight are independent. They can be chosen as
\begin{equation}
[I^{03}, K^{03}] = 0, \qquad [I^{12}, K^{03}] = 0, 
\qquad [I^{01}, K^{31}] =  K^{03}
\end{equation}
and
$$[I^{01}, K^{03}] =  K^{31}, \qquad [I^{02}, K^{03}] = - K^{23}, 
\qquad [I^{31}, K^{03}] =  K^{01},$$
\begin{equation}
[I^{23}, K^{03}] = - K^{02}, \qquad [I^{23}, K^{31}] = K^{12}.
\end{equation}
The other 28 conditions in (16) are consequences of (17) and (18) and the commutation relations for the $I^{\mu\nu}$. With the operator $K^{31}$ expressed in terms of $K^{03}$ from the first equality in (18) and substituted in (17), we
obtain three independent equations that must be satisfied by $K^{03}$. Finding $K^{03}$ and using formulas (18), we obtain all the components of the antisymmetric operator $K^{\mu\nu}$.

Solving system of equations (17) for the unknown operator $K^{03}$ means finding the result of the action of this operator on vectors of the canonical basis 
$\xi_{(l_{0}, l_{1})lm}$ in any irreducible representation $(l_{0}, l_{1})$ of the proper Lorentz group. Using the known action of $I^{12}$, $I^{03}$, and 
$I^{01}$ on basis vectors $\xi_{(l_{0}, l_{1})lm}$, we obtain
\begin{eqnarray}
&{}&K^{03} \xi_{(l_{0}, l_{1})lm} = f(l_{0}-1, l_{1}-1; l_{0}, l_{1}) 
\left[ a(l,m) \sqrt{(l+l_{0}-1)(l+l_{0})(l+l_{1}-1)(l+l_{1})} \times \right. \nonumber \\
& &\times \xi_{(l_{0}-1, l_{1}-1)l-1 m} - b(l,m) 
\sqrt{(l-l_{0}+1)(l+l_{0})(l-l_{1}+1)(l+l_{1})} 
\xi_{(l_{0}-1, l_{1}-1)l m} - \nonumber \\
& &\left. - a(l+1,m) \sqrt{(l-l_{0}+1)(l-l_{0}+2)(l-l_{1}+1)(l-l_{1}+2)} 
\xi_{(l_{0}-1, l_{1}-1)l+1 m} \right] + \nonumber \\
& &+ f(l_{0}-1, l_{1}+1; l_{0}, l_{1})
\left[ a(l,m) \sqrt{(l+l_{0}-1)(l+l_{0})(l-l_{1}-1)(l-l_{1})} 
\xi_{(l_{0}-1, l_{1}+1)l-1 m} + \right. \nonumber \\ 
& &+ b(l,m) \sqrt{(l-l_{0}+1)(l+l_{0})(l-l_{1})(l+l_{1}+1)} 
\xi_{(l_{0}-1, l_{1}+1)l m} - \nonumber \\
& &\left. - a(l+1,m) \sqrt{(l-l_{0}+1)(l-l_{0}+2)(l+l_{1}+1)(l+l_{1}+2)}  
\xi_{(l_{0}-1, l_{1}+1)l+1 m} \right] + \nonumber \\
& &+ f(l_{0}+1, l_{1}-1; l_{0}, l_{1})
\left[ a(l,m) \sqrt{(l-l_{0}-1)(l-l_{0})(l+l_{1}-1)(l+l_{1})} 
\xi_{(l_{0}+1, l_{1}-1)l-1 m} + \right. \nonumber \\ 
& &+ b(l,m) \sqrt{(l-l_{0})(l+l_{0}+1)(l-l_{1}+1)(l+l_{1})} 
\xi_{(l_{0}+1, l_{1}-1)l m} - \nonumber \\
& &\left. - a(l+1,m) \sqrt{(l+l_{0}+1)(l+l_{0}+2)(l-l_{1}+1)(l-l_{1}+2)}  
\xi_{(l_{0}+1, l_{1}-1)l+1 m} \right] + \nonumber \\
& &+ f(l_{0}+1, l_{1}+1; l_{0}, l_{1})
\left[ a(l,m) \sqrt{(l-l_{0}-1)(l-l_{0})(l-l_{1}-1)(l-l_{1})} 
\xi_{(l_{0}+1, l_{1}+1)l-1 m} - \right. \nonumber \\ 
& &- b(l,m) \sqrt{(l-l_{0})(l+l_{0}+1)(l-l_{1})(l+l_{1}+1)} 
\xi_{(l_{0}+1, l_{1}+1)l m} - \nonumber \\
& &\left. - a(l+1,m) \sqrt{(l+l_{0}+1)(l+l_{0}+2)(l+l_{1}+1)(l+l_{1}+2)}  
\xi_{(l_{0}+1, l_{1}+1)l+1 m} \right] + \nonumber \\
& &+ f_{1}(l_{0}, l_{1}; l_{0}, l_{1})
\left[ a(l,m) \sqrt{(l-l_{0})(l+l_{0})(l-l_{1})(l+l_{1})} 
\xi_{(l_{0}, l_{1})l-1 m}  
- b(l,m) l_{0} l_{1} \xi_{(l_{0}, l_{1})l m} - \right. \nonumber \\
& &\left. - a(l+1,m) \sqrt{(l-l_{0}+1)(l+l_{0}+1)(l-l_{1}+1)(l+l_{1}+1)}  
\xi_{(l_{0}, l_{1})l+1 m} \right] + \nonumber \\
& &+ f_{2}(l_{0}, l_{1}; l_{0}, l_{1}) m \xi_{(l_{0}, l_{1})l m},
\end{eqnarray}
where
\begin{equation}
a(l, m) = \frac{1}{l}\sqrt{\frac{(l-m)(l+m)}{4l^{2}-1}}, \qquad
b(l, m) = \frac{m}{l(l+1)},
\end{equation}
and $f(l_{0} \pm 1, l_{1}-1; l_{0}, l_{1})$, 
$f(l_{0} \pm 1, l_{1} + 1; l_{0}, l_{1})$, and 
$f_{j}(l_{0},l_{1}; l_{0},l_{1})$, $j = 1,2$, are arbitrary constants.

In the family of antisymmetric tensor operators $K^{\mu\nu}$, we can single out two types of operators that transform under nonequivalent irreducible representations of the orthochronous Lorentz group. We let $\Gamma^{\mu\nu}$
and $L^{\mu\nu}$ denote them, assuming that contractions of these operators with two arbitrary polar four-vectors respectively give scalar and pseudoscalar operators of the $L^{\uparrow}$ group. This means that the transformation
properties of antisymmetric operators of these two types under the spatial reflection $P$ are described by the relations
\begin{equation}
P K^{0i} + \eta K^{0i} P = 0, \qquad P K^{jk}- \eta K^{jk} P = 0, \qquad
 i,j,k = 1,2,3,
\end{equation}
where $\eta = +1$ if $K^{\mu\nu} = \Gamma^{\mu\nu}$ and $\eta = -1$ if 
$K^{\mu\nu} = L^{\mu\nu}$.

Because (see [7], [8])
\begin{equation}
P \xi_{(l_{0}, l_{1})l m} = \pm (-1)^{[l]} \xi_{(-l_{0}, l_{1})l m},
\end{equation}
where the plus or minus sign is taken the same for all irreducible 
$L^{\uparrow}_{+}$ representations belonging to a given representation $S_{0}$, we can use relations (19)--(21) to obtain the constraints relating the constants of $\Gamma^{\mu\nu}$ and $L^{\mu\nu}$:
\begin{eqnarray}
& &f(l_{0} \pm 1, l_{1}-1; l_{0}, l_{1}) =
\eta f(-l_{0} \mp 1, l_{1}-1; -l_{0}, l_{1}), \nonumber \\
& &f(l_{0} \pm 1, l_{1}+1; l_{0}, l_{1}) =
\eta f(-l_{0} \mp 1, l_{1}+1; -l_{0}, l_{1}), \nonumber \\
& &f_{1}(l_{0}, l_{1}; l_{0}, l_{1}) = \eta f_{1}(-l_{0}, l_{1}; -l_{0}, l_{1}), \nonumber \\
& &f_{2}(l_{0}, l_{1}; l_{0}, l_{1}) = 
-\eta f_{2}(-l_{0}, l_{1}; -l_{0}, l_{1}). 
\end{eqnarray}

Electromagnetic current (5), (6) involves a second-rank antisymmetric tensor operator of the $\Gamma^{\mu\nu}$ type.

\begin{center}
{\large \bf 4. Antisymmetric tensor operators in the theory \\ of ISFIR-class
fields with double symmetry}
\end{center}

To eliminate the infinite arbitrariness in the constants of the operators  
$\Gamma^{\mu\nu}$ in considering electromagnetic current (5), (6) within the theory of ISFIR-class fields, we assume that this current is invariant under secondary symmetry transformations (4). Hence, the conditions
\begin{equation}
[D^{\alpha}, \Gamma^{\mu\nu}] = 0.
\end{equation}
must be satisfied. Among the 24 relations in (24), there are only two that are independent,
\begin{equation}
[D^{0}, \Gamma^{03}] = 0, \qquad [D^{0}, \Gamma^{12}] = 0.
\end{equation}
and the other 22 conditions in (24) are consequences of these two and also of
relations (16) and the commutation relations of the four-vector operator with
infinitesimal operators of the $L^{\uparrow}_{+}$ group (see [7]--[9]).

From the countable set of versions of the doubly symmetric theory of ISFIR-class fields with spontaneously broken secondary symmetry [9], [13], in what follows, we choose to work with the simplest one characterized by a physically acceptable mass spectrum [14]. In the chosen version of the theory, first, the field transforms under the proper Lorentz group representation consisting of all those and only those irreducible finite-dimensional representations that contain spin 1/2:
\begin{equation}
S^{3/2} = \bigoplus_{N=1}^{+\infty} \left[ \left( -\frac{1}{2}, \frac{1}{2}+N
\right) \oplus \left( \frac{1}{2}, \frac{1}{2}+N \right) \right].
\end{equation}
Second, the transformations of secondary symmetry (4) form a four-parameter Abelian group: $[D^{\mu}, D^{\nu}] = 0$. Third, the polar four-vector operators $D^{\mu}$ from transformations (4) and  $\Gamma^{\mu}$ from Eq. (7) coincide with each other up to a constant numerical factor, and
\begin{eqnarray}
& & d_{0}^{-1} D^{0}  \xi_{(\pm \frac{1}{2}, l_{1})lm} =
c_{0}^{-1} \Gamma^{0} \xi_{(\pm \frac{1}{2}, l_{1})lm} = 
\frac{l+1/2}{l_{1}^{2}-1/4} \xi_{(\mp \frac{1}{2}, l_{1})lm} - \nonumber \\
& & -\frac{\sqrt{(l_{1}-l-1)(l_{1}+l)}}{l_{1} - 1/2}
\xi_{(\pm \frac{1}{2}, l_{1}-1)lm} -
\frac{\sqrt{(l_{1}-l)(l_{1}+l+1)}}{l_{1} + 1/2}
\xi_{(\pm \frac{1}{2}, l_{1}+1)lm},
\end{eqnarray}
where $d_{0}$ and $c_{0}$ are arbitrary real constants. Fourth, there exist only two four-vector operators (up to a numerical factor) whose components commute with the components of $D^{\mu}$: the operator $D^{\mu}$ itself and the axial four-vector operator $L^{\mu}$, whose time component acts on the canonical basis vectors as
\begin{eqnarray}
& & b_{0}^{-1} L^{0}  \xi_{(\pm \frac{1}{2}, l_{1})lm} =
\mp \frac{l_{1}(l+1/2)}{l_{1}^{2}-1/4} \xi_{(\mp \frac{1}{2}, l_{1})lm} \pm \nonumber \\
& & \pm \frac{\sqrt{(l_{1}-l-1)(l_{1}+l)}}{2 l_{1} - 1}
\xi_{(\pm \frac{1}{2}, l_{1}-1)lm} \mp
\frac{\sqrt{(l_{1}-l)(l_{1}+l+1)}}{2l_{1} + 1}
\xi_{(\pm \frac{1}{2}, l_{1}+1)lm},
\end{eqnarray}
where $b_{0}$ is an arbitrary constant.

Direct calculations show that there exists only one antisymmetric tensor operator  $\Gamma^{\mu\nu}$ (up to a numerical factor) that satisfies conditions (25) with relations (19), (20), (23) (with  $\eta = 1$), and (27) taken into
account. The values of its constants are
\begin{eqnarray}
& &f\left( \pm \frac{1}{2}, l_{1}+1; \mp \frac{1}{2}, l_{1} \right) =
f\left( \pm \frac{1}{2}, l_{1}; \mp \frac{1}{2}, l_{1}+1 \right) = 
\frac{ih_{0}}{l_{1}+1/2},
\nonumber \\
& &f_{1}\left( \pm \frac{1}{2}, l_{1}; \pm \frac{1}{2}, l_{1} \right) =
\frac{h_{0}}{l_{1}^{2} - 1/4}, 
\nonumber \\
& &f_{2}\left( \pm \frac{1}{2}, l_{1}; \pm \frac{1}{2}, l_{1} \right) =
\mp \frac{2h_{0}l_{1}}{l_{1}^{2} - 1/4},
\end{eqnarray}
where $h_{0}$  is an arbitrary number. Such an operator $\Gamma^{\mu\nu}$ can be expressed in terms of the four-vector operators $\Gamma^{\mu}$ and $L^{\mu}$ described by formulas (27) and (28):
$\Gamma^{\mu}$ п╦ $L^{\mu}$:
\begin{equation}
\Gamma^{\mu\nu} = b_{0}^{-2}h_{0} (L^{\mu}L^{\nu}-L^{\nu}L^{\mu})
\end{equation}
or
\begin{equation}
\Gamma^{\mu\nu}= ib_{0}^{-1}c_{0}^{-1}h_{0}\varepsilon^{\mu\nu\rho\sigma}
\Gamma_{\rho}L_{\sigma}.
\end{equation}

Finding the general structure of the matrix tensor operators 
$\Gamma^{\mu\nu\nu_{1}\ldots \nu_{j}}$, $j=1,2,\ldots$, of rank three and
higher that are antisymmetric in two indices and lead to polar current four-vector (5) is a sufficiently laborious but feasible problem. Knowing the general structure would allow finding all the sets of the constants of such operators that ensure the invariance of current (5), (6) under secondary symmetry transformations (4). Lacking this information, we ensure the double invariance of the theory by considering only those higher-rank operators 
$\Gamma^{\mu\nu\nu_{1}\ldots \nu_{j}}$ that are expressed in terms of the 
four-vector operators $\Gamma^{\mu}$ in (27) and $L^{\mu}$ in (28). It is quite possible that this exhausts the list of admissible operators 
$\Gamma^{\mu\nu\nu_{1}\ldots \nu_{j}}$, but we have no proof of
this.

The tracelessness that we require of the rank-($2+j$) matrix operator
$\Gamma^{\mu\nu\nu_{1}\ldots \nu_{j}}$ implies that this operator
does not contain matrix tensor operators of a lower rank $j$. In expressing traceless operators $\Gamma^{\mu\nu\nu_{1}\ldots \nu_{j}}$ in terms of the operators $\Gamma^{\mu}$ and $L^{\mu}$, we must use the relations
\begin{eqnarray}
& &\Gamma^{\mu} \Gamma_{\mu} = 4c_{0}^{2}, \qquad L^{\mu} L_{\mu} = -3b_{0}^{2},\qquad L^{\mu} \Gamma_{\mu} =0, \\
& &4b_{0}^{-2}(L^{\mu} L^{\nu}+L^{\nu} L^{\mu}) + 6 g^{\mu\nu} =
c_{0}^{-2}(\Gamma^{\mu} \Gamma^{\mu}+\Gamma^{\nu} \Gamma^{\mu}) - 2 g^{\mu\nu}.
\end{eqnarray}
The last equality reflects the coincidence (up to a numerical factor) of the second-rank traceless symmetric tensor operators constructed from the operators $L^{\mu}$ and $\Gamma^{\mu}$.

We now list all the linearly independent matrix operators of rank three and four with the required properties:
\begin{eqnarray}
& &\Gamma^{\mu\nu\rho}_{1} = h_{1}\Gamma^{\rho}\Gamma^{\mu\nu},  
\nonumber \\
& &\Gamma^{\mu\nu\rho}_{2} = h_{2}[L^{\rho}\tilde{\Gamma}^{\mu\nu}-
\frac{b_{0}h_{0}}{c_{0}}(g^{\mu\rho}\Gamma^{\nu}-g^{\nu\rho}\Gamma^{\mu})],
\nonumber \\
& &\Gamma^{\mu\nu\rho\sigma}_{1} = h_{3}[\Gamma^{\rho}\Gamma^{\sigma}
\Gamma^{\mu\nu} -\frac{c_{0}^{2}}{3}(4g^{\rho\sigma}
\Gamma^{\mu\nu}+g^{\mu\rho}\Gamma^{\nu\sigma}-g^{\nu\rho}\Gamma^{\mu\sigma}
+g^{\mu\sigma}\Gamma^{\nu\rho}-g^{\nu\sigma}\Gamma^{\mu\rho})], 
\nonumber \\
& &\Gamma^{\mu\nu\rho\sigma}_{2} = h_{4}[L^{\rho}\Gamma^{\sigma}
\tilde{\Gamma}^{\mu\nu}-\frac{b_{0}h_{0}}{c_{0}}(g^{\mu\rho}\Gamma^{\nu}
\Gamma^{\sigma}-g^{\nu\rho}\Gamma^{\mu}\Gamma^{\sigma})+  \nonumber \\
& &\qquad +\frac{c_{0}b_{0}}{6}(2g^{\rho\sigma}\Gamma^{\mu\nu}-
g^{\mu\rho}\Gamma^{\nu\sigma}+g^{\nu\rho}\Gamma^{\mu\sigma}+
5g^{\mu\sigma}\Gamma^{\nu\rho}-5g^{\nu\sigma}\Gamma^{\mu\rho})],
\end{eqnarray}
where
\begin{equation}
\tilde{\Gamma}^{\mu\nu} = \frac{i}{2}\varepsilon^{\mu\nu\rho\sigma}
\Gamma_{\rho\sigma} = \frac{h_{0}}{b_{0}c_{0}}(\Gamma^{\mu}L^{\nu}
-\Gamma^{\nu}L^{\mu}),
\end{equation}
and $h_{1}$, $h_{2}$, $h_{3}$, and $h_{4}$ are arbitrary constants.

\begin{center}
{\large {\bf 5. Finite proper Lorentz transformations for the finite-dimensional
irreducible representations} $(\pm 1/2, l_{1})$}
\end{center}

We consider some consequences of the proper Lorentz transformation corresponding to the transition from the particle rest frame to the laboratory frame in which the particle moves along the third coordinate axis with a velocity $v$. The corresponding transformations in the space of wave vectors $\psi$ are 
implemented by the operator $S(\alpha) = e^{\alpha I^{03}}$, where 
$\tanh \alpha = v$. Because the action of the infinitesimal operator $I^{03}$
on a vector of the canonical basis does not change its spin projection on the third axis, the matrix elements of $I^{03}$ and the matrix elements of finite proper Lorentz transformations in the space of any irreducible 
$L^{\uparrow}_{+}$ representation $(l_{0},l_{1})$ are determined by the 
relations
\begin{equation}
I^{03} \xi_{(l_{0}, l_{1})l m} = \sum_{l'} I^{(l_{0},l_{1})}_{l'm,lm}
\xi_{(l_{0}, l_{1})l' m},
\end{equation}
\begin{equation}
e^{\alpha I^{03}} \xi_{(l_{0}, l_{1})l m} = 
\sum_{l'} A^{(l_{0},l_{1})}_{l'm,lm} (\alpha)
\xi_{(l_{0}, l_{1})l' m}.
\end{equation}

We establish a number of constraints relating the quantities 
$A^{(l_{0},l_{1})}_{l'm,lm} (\alpha)$, restricting ourself to finite-dimensional irreducible representations of the $L^{\uparrow}_{+}$ group. It follows from the explicit form of the decomposition coefficients in (36) for such representations (see [7]--[9]) that
\begin{equation}
I^{(l_{0},l_{1})}_{l'-m,l-m} = I^{(-l_{0},l_{1})}_{l'm,lm} = 
-\left[ I^{(l_{0},l_{1})}_{l'm,lm} \right]^{*}.
\end{equation}
From this and relations (36) and (37), we conclude that
\begin{equation}
A^{(l_{0},l_{1})}_{l'-m,l-m} (\alpha) = A^{(-l_{0},l_{1})}_{l'm,lm} (\alpha) = 
\left[ A^{(l_{0},l_{1})}_{l'm,lm} (-\alpha) \right]^{*}.
\end{equation}

We note that wave vectors of the form [10]
\begin{equation}
\psi_{m}(p_{0}) = \sum_{N=1}^{+\infty} \chi (N) \left[ 
\xi_{(\frac{1}{2},\frac{1}{2}+N)\frac{1}{2} m} + 
r \xi_{(-\frac{1}{2},\frac{1}{2}+N)\frac{1}{2} m} \right], 
\end{equation}
correspond to two independent states of the rest spin-1/2 particle that satisfy Eq. (7), where $r$ is equal to 1 or -1 and describes the spatial parity of the state and $m=-1/2, 1/2$. In the reference frame where the particle velocity is directed along the third axis, the wave vector of the particle is given by
\begin{eqnarray}
\psi_{m}(p) = S(\alpha)\psi_{m}(p_{0})&=& \sum_{N=1}^{+\infty} 
\sum_{l=1/2}^{N-1/2}
\chi (N) \left[ A^{(\frac{1}{2}, \frac{1}{2}+N)}_{lm,\frac{1}{2}m} (\alpha)
\xi_{(\frac{1}{2},\frac{1}{2}+N)\frac{1}{2} m} + \right. \nonumber \\
& &\left. + r A^{(-\frac{1}{2}, \frac{1}{2}+N)}_{lm,\frac{1}{2}m} (\alpha)
\xi_{(-\frac{1}{2},\frac{1}{2}+N)\frac{1}{2} m} \right].  
\end{eqnarray} 
Taking relations (39) into account, we hence conclude that to calculate the form factors in (8) and (9), we need only know the quantities 
$A^{(\frac{1}{2},l_{1})}_{l\frac{1}{2},\frac{1}{2}\frac{1}{2}} (\alpha)$ and only with $l=1/2$ and $3/2$ if we restrict ourself to tensor
operators of the second, third, and fourth ranks in formula (6).

Taking the derivative of both sides of (37) with respect to $\alpha$ and using the known form of the operator $I^{03}$ in the canonical basis (see [7]--[9]), we obtain
\begin{eqnarray}
\frac{d}{d\alpha} 
A^{(\frac{1}{2},l_{1})}_{l\frac{1}{2},\frac{1}{2}\frac{1}{2}} (\alpha) &=& 
-\frac{i}{4(l+1)}\sqrt{(2l+1)(2l+3)(l_{1}^{2}-(l+1)^{2})}
A^{(\frac{1}{2},l_{1})}_{l+1\frac{1}{2},\frac{1}{2}\frac{1}{2}} (\alpha)+
\nonumber \\
& &+\frac{l_{1}}{l(l+1)} 
A^{(\frac{1}{2},l_{1})}_{l\frac{1}{2},\frac{1}{2}\frac{1}{2}} (\alpha)+
\frac{i}{4l}\sqrt{(2l-1)(2l+1)(l_{1}^{2}-l^{2})}
A^{(\frac{1}{2},l_{1})}_{l-1\frac{1}{2},\frac{1}{2}\frac{1}{2}} (\alpha).
\end{eqnarray}
For a given number $l_{1}$, starting with a single matrix element 
$A^{(\frac{1}{2},l_{1})}_{\frac{1}{2}\frac{1}{2},\frac{1}{2}\frac{1}{2}} 
(\alpha)$ as a function of $\alpha$, this recurrence relation allows finding all the other elements 
$A^{(\frac{1}{2},l_{1})}_{l\frac{1}{2},\frac{1}{2}\frac{1}{2}} (\alpha)$ with 
$l = 3/2, \ldots, |l_{1}|-1$.

We now derive a recurrence relation involving the 
$A^{(\frac{1}{2},l_{1})}_{l\frac{1}{2},\frac{1}{2}\frac{1}{2}}(\alpha)$ with different values of $l_{1}$. In the representation space $(1/2, l_{1}) \oplus (1/2, l_{1}+1)$ of the proper Lorentz group, we take two vectors $\varphi(p)$ and $\Phi(p)$ that are related to each other by the relativistically covariant formula
\begin{equation}
\Phi(p) = (L^{\mu}p_{\mu}) \varphi(p)
\end{equation}
in an arbitrary inertial reference frame, where the four-momentum of the particle is equal to $p$, and have only one nonvanishing component each in the canonical basis in the particle rest frame,
\begin{equation}
\varphi(p_{0}) = \varphi_{0} \xi_{(\frac{1}{2}, l_{1})\frac{1}{2} \frac{1}{2}}, 
\qquad
\Phi(p_{0}) = \Phi_{0} \xi_{(\frac{1}{2}, l_{1}+1)\frac{1}{2} \frac{1}{2}},
\end{equation}
where $\varphi_{0}$ and $\Phi_{0}$ are arbitrary constants. The full correspondence between the general structure of the four-vector operator, equality (43), which in the particle rest frame has the form
\begin{equation}
\Phi(p_{0}) = (L^{0}M) \varphi(p_{0}),
\end{equation}
and relations (44) is attained under the conditions
\begin{eqnarray}
& &L^{0} \xi_{(\frac{1}{2}, l_{1})\frac{1}{2} \frac{1}{2}}= 
i a_{0} \sqrt{(l_{1}-1/2)(l_{1}+3/2)}
\xi_{(\frac{1}{2}, l_{1}+1)\frac{1}{2} \frac{1}{2}},
\nonumber \\
& &\Phi_{0} = i \varphi_{0} a_{0}M \sqrt{(l_{1}-1/2)(l_{1}+3/2)}.
\end{eqnarray}

In the laboratory frame, which is characterized by the Lorentz boost parameter 
$\alpha$, it follows from (43) with formulas (37) and (44) taken into account that
\begin{equation}
\sum_{l} \Phi_{0} 
A^{(\frac{1}{2},l_{1}+1)}_{l\frac{1}{2},\frac{1}{2}\frac{1}{2}} (\alpha)
\xi_{(\frac{1}{2}, l_{1}+1)l \frac{1}{2}} = (L^{0}M \cosh \alpha - 
L^{3} M \sinh \alpha) \sum_{l'} \phi_{0}
A^{(\frac{1}{2},l_{1})}_{l'\frac{1}{2},\frac{1}{2}\frac{1}{2}} (\alpha)
\xi_{(\frac{1}{2}, l_{1})l' \frac{1}{2}}.
\end{equation}
Because $L^{3} = [L^{0}, I^{03}]$ (see [7]--[9]), using the explicit form of 
$I^{03}$in the canonical basis and formulas (46), we obtain the recurrence relation in $l_{1} = 3/2, 5/2, \ldots$ with $l = 1/2, \ldots, l_{1}$
\begin{eqnarray}
A^{(\frac{1}{2},l_{1}+1)}_{l\frac{1}{2},\frac{1}{2}\frac{1}{2}} (\alpha) &=&
\frac{1}{2l(l+1)\sqrt{(2l_{1}-1)(2l_{1}+3)}} \times \nonumber \\
& &\times \left[-il\sinh \alpha \sqrt{(2l+1)(2l+3)(l_{1}-l-1)(l_{1}-l)}
A^{(\frac{1}{2},l_{1})}_{l+1\frac{1}{2},\frac{1}{2}\frac{1}{2}} (\alpha)+
\right. \nonumber \\
& &+(4l(l+1)\cosh \alpha +\sinh \alpha)\sqrt{(l_{1}-l)(l_{1}+l+1)}
A^{(\frac{1}{2},l_{1})}_{l\frac{1}{2},\frac{1}{2}\frac{1}{2}} (\alpha)+
\nonumber \\
& &\left. +i(l+1)\sinh \alpha \sqrt{(2l-1)(2l+1)(l_{1}+l)(l_{1}+l+1)}
A^{(\frac{1}{2},l_{1})}_{l-1\frac{1}{2},\frac{1}{2}\frac{1}{2}} (\alpha)
\right] ,
\end{eqnarray}
from (47).

We can now prove by induction that the equality
\begin{equation}
A^{(\frac{1}{2},l_{1})}_{\frac{1}{2} \frac{1}{2},\frac{1}{2}\frac{1}{2}} (\alpha) = \frac{2}{l_{1}^{2}-1/4} \sum_{n=0}^{l_{1}-3/2} (l_{1}-n-1/2) 
e^{(l_{1}-2n-1)\alpha} ,
\end{equation}
where $l_{1} = 3/2, 5/2, \ldots$, holds. First, we verify formula (49) for 
$l_{1} = 3/2$, when it becomes
\begin{equation}
A^{(\frac{1}{2},\frac{3}{2})}_{\frac{1}{2} \frac{1}{2},\frac{1}{2}\frac{1}{2}} (\alpha) = e^{\alpha /2}.
\end{equation}
For this, it suffices to use the known equality [7]--[9]
\begin{equation}
I^{03} \xi_{(\frac{1}{2},\frac{3}{2}) \frac{1}{2} \frac{1}{2}} = \frac{1}{2}
\xi_{(\frac{1}{2},\frac{3}{2}) \frac{1}{2} \frac{1}{2}}
\end{equation}
and relation (37). Assuming that formula (49) holds for some value of $l_{1}$ equal to $k_{1}$, $k_{1} = 5/2, 7/2, \ldots$, we use (42) to obtain
\begin{equation}
A^{(\frac{1}{2},k_{1})}_{\frac{3}{2} \frac{1}{2},\frac{1}{2}\frac{1}{2}}
(\alpha) = \frac{2i\sqrt{2}}{(k_{1}^{2}-1/4)\sqrt{k_{1}^{2}-9/4}}
\sum_{n=0}^{l_{1}-3/2} (k_{1}-n-1/2) (k_{1}-3n-3/2) e^{(k_{1}-2n-1)\alpha}.
\end{equation}
Using recurrence relation (48), we now find that the quantity 
$A^{(\frac{1}{2},k_{1}+1)}_{\frac{1}{2} \frac{1}{2},\frac{1}{2}\frac{1}{2}}(\alpha)$ obtained from it equals the right-hand side of (49) if $l_{1}$ there is replaced with $k_{1}+1$, in other words, that formula (49) also holds for
$l_{1} = k_{1}+1$.

\begin{center}
{\large \bf 6. Writing the field vector components as continued fractions}
\end{center}

We address the question of calculating the field vector $\psi (p)$ belonging to the $L^{\uparrow}_{+}$-representa- tion space $S^{3/2}$ in (26) and satisfying Eq. (7), where the operator $\Gamma^{0}$ is given by relation (27) and the operator 
$R$ has the form [13], [14]
\begin{equation}
R \xi_{(\pm \frac{1}{2}, \frac{1}{2}+N)lm} = 
\rho(N) \xi_{(\pm \frac{1}{2}, \frac{1}{2}+N)lm} =
\left[ \sum_{i} \lambda_{i} \eta_{i}(N) \right] 
\xi_{(\pm \frac{1}{2}, \frac{1}{2}+N)lm},
\end{equation}
with
\begin{equation}
\eta_{i}(N) = 2\frac{v_{i}^{N}(v_{i}N+N+1)-w_{i}^{N}(w_{i}N+N+1)}
{N(N+1)(v_{i}-w_{i})(2+v_{i}+w_{i})},
\end{equation}
where $N = 1, 2, \ldots$, $v_{i}=(z_{i}+\sqrt{z_{i}^{2}-4})/2$, 
$w_{i}= (z_{i}-\sqrt{z_{i}^{2}-4})/2$, $\lambda_{i}$ and $z_{i}$ are some real constants, and the index $i$ takes one or several values.

We introduce a reference frame where the four-momentum is equal to 
$p_{0} = \{ M_{0}, 0, 0, 0 \}$ and consider the state of the field 
$\psi (p_{0})$ that has the spin 1/2 in this frame and is described by (40). Then Eq. (7) reduces to the recurrence relation for the components of the field vector $\chi (N)$
\begin{eqnarray}
& &N\sqrt{N(N+2)} \chi (N+1) + (N+1)\sqrt{(N-1)(N+1)} \chi (N-1)+ 
\nonumber \\
& &\qquad +[(M_{0}c_{0})^{-1} N(N+1) \rho(N) - r] \chi (N) = 0,
\end{eqnarray}
where $N = 1, 2, \ldots$ and also $\chi (0)=0$.

For any values of free parameters of the theory, relation (55) allows expressing all the $\chi (N)$, $N = 2, 3, \ldots$, in terms of a single quantity 
$\chi (1)$. Only those sets of free parameters of the theory for which the obtained $\chi (N)$ ensure normalizability of the corresponding field vectors 
$\psi (p_{0})$ in (40) are interesting in particle physics. Satisfaction of this requirement for one or another fixed set of free parameters $z_{i}$, 
$\lambda_{i}$, and $c_{0}$ leads to a set of values of $M_{0}$ being selected and acquiring the status of the masses of particles with rest spin 1/2 and spatial parity $r$.

A detailed analysis of the characteristics of the mass spectrum of the theory depending on the domain of the free parameters $z_{i}$ was given in [14]. It was shown that the experimental picture of nucleon resonances satisfactorily agrees with the mass spectrum of the theory of ISFIR-class fields under study if 
$z_{1} > 2$ and $|z_{j}| < 2$ for $j \neq 1$ is assumed in formulas (53) and (54). In this case, which we consider in what follows, the asymptotic form of the components of field vectors is determined by a single parameter $z_{1}$: as $N \rightarrow \infty$, we have
\begin{equation}
\chi(N) = A_{0} G(N) (1+{\cal O}(N^{-1})) 
+ B_{0} G^{-1}(N) (1+{\cal O}(N^{-1})),
\end{equation}
where
\begin{equation}
G(N) = \left( -\frac{\rho(1)}{M_{0}c_{0}} \right)^{N} 
\frac{v_{1}^{N(N+1)/2}}{N!}.
\end{equation}

The field vector $\psi (p_{0})$ in (40) is normalizable if and only if the coefficient $A_{0}$ in (56), which depends on the free parameters of the theory $M_{0}$, $c_{0}$, and $\lambda_{i}$, vanishes. We have no analytic methods for solving the equations $A_{0} = 0$ for the parameter $M_{0}$, i.e., methods for finding the mass spectrum inherent to Eq. (55). The problem of calculating any number of the lower values of mass is solved approximately via numerical methods whose algorithm was described in [14]. But any approximation to the mass value leads to a nonzero, although small, quantity $A_{0}$, and the quantities 
$\chi(N)$ found from exact relation (55) hence change their behavior from a decrease to a rapid increase as $N$ increases already for small numbers $N$. 
(For example, for the parameter values of the theory given at the beginning of Sec. 8, the sequence $\chi(N)$ starts increasing at the eighth term.) This considerably reduces the possible use of numerical computations.

In what follows, we give a formula for calculating $\chi(N)$ with the following properties. In and of itself, this formula is an approximate relation for 
$\chi(N)$ if the parameter $M_{0}$ is different from the state mass $M$ and an exact relation if $M_{0} = M$. The values of $\chi(N)$ obtained from the formula tend to zero as $N$ increases infinitely.

We introduce the notation
\begin{equation}
\pi(N) = (-1)^{N} \sqrt{N(N+1)} \chi(N), \qquad 
H(N) = (M_{0}c_{0})^{-1} N(N+1) \rho(N) - r .
\end{equation}
Relation (55) can then be rewritten as
\begin{equation}
N^{2} \pi(N+1) + (N+1)^{2} \pi(N-1) - H(N) \pi(N) = 0,
\end{equation}
where $N = 1, 2, \ldots$. This system must be supplemented by the condition 
$\pi(0) = 0$ or the equivalent condition
\begin{equation}
\pi(2) = H(1) \pi(1)
\end{equation}
obtained from (59) with $N = 1$. Equations (59) together with (60) allow expressing all the required quantities $\pi(N)$ in terms of $\pi(1)$. We have
\begin{equation}
\frac{\pi(N)}{\pi(N+1)} = \frac{N^{2}}{\displaystyle H(N) - 
\frac{(N-1)^{2} (N+1)^{2}}{\displaystyle H(N-1) - \frac{(N-2)^{2}N^{2}}
{\displaystyle \ddots -\frac{\ddots}{\displaystyle H(2) - 
\frac{9}{\displaystyle H(1)}}}}} ,
\end{equation}
where $N = 2, 3, \ldots$.

We now suppose that the satisfaction of the condition $\pi(0) = 0$ is not a necessary addition to relation (59). We therefore drop one equation in (59), the one with $N = 1$, from consideration. The remaining equations of system (59) (those with $N = 2, 3, \ldots$) have two independent solutions because all the quantities $\pi(N)$ can be expressed in terms of $\pi(1)$ and $\pi(2)$. One of these solutions is given by (60) and (61), and the other can be represented as the infinite continued fraction
\begin{equation}
\frac{\pi(N+1)}{\pi(N)} = \frac{(N+2)^{2}}{\displaystyle H(N+1) - 
\frac{(N+1)^{2} (N+3)^{2}}{\displaystyle H(N+2) - \frac{(N+2)^{2}(N+4)^{2}}
{\displaystyle \ddots -\frac{\ddots}{\displaystyle H(N+n) - 
\frac{(N+n)^{2}(N+n+2)^{2}}{\displaystyle \ddots}}}}} ,
\end{equation}
where $N = 1, 2, \ldots$. This infinite continued fraction must be regarded as the limit of the sequence
\begin{equation}
\left[ \frac{\pi(N+1)}{\pi(N)} \right]_{1} = \frac{(N+2)^{2}}{H(N+1)}, \qquad
\left[ \frac{\pi(N+1)}{\pi(N)} \right]_{2} = \frac{(N+2)^{2}}{\displaystyle 
H(N+1) -  \frac{(N+1)^{2} (N+3)^{2}}{\displaystyle H(N+2)}}, \; \; \ldots ,
\end{equation}
which emerges as a result of solving system of equations (59) with 
$N = 2, 3, \ldots$ by consecutive approximations:
\begin{equation}
N^{2} \left[ \frac{\pi(N+1)}{\pi(N)} \right]_{n-1} + (N+1)^{2} 
\left\{ \left[ \frac{\pi(N)}{\pi(N-1)} \right]_{n} \right\}^{-1} - H(N) = 0,
\end{equation}
where $n = 1, 2, \ldots$, with $[\pi(N+1)/ \pi(N)]_{0} = 0$.

We prove that in the considered domain of the parameters $z_{i}$ in formulas (53) and (54), the sequence $[\pi(N+1)/\pi(N)]_{n}$ converges as 
$n \rightarrow \infty$. For this, we recall that $\eta_{1} (N)$ increases exponentially as $N$ increases, and the values of $\eta_{j} (N)$, $j \neq 1$, oscillate in a bounded region as $N$ varies. Therefore, for sufficiently large
values of $N$, the contribution of the $\eta_{j} (N)$, $j \neq 1$, to $\rho (N)$ and then to $H(N)$ becomes arbitrarily small. Hence, for any set of the free parameters $M_{0}$, $c_{0}$, and $\lambda^{i}$ of the theory and for any number  $\varepsilon > 0$, a number $N_{0}$ depending on them can be given such that for all $N \ge N_{0}$, the inequality $0 < N^{4}/H(N) < \varepsilon$  holds (for
simplicity in what follows, we assume that $M_{0}c_{0}\rho(N_{0}) > 0$), and hence so does the inequality
\begin{equation}
0 < \frac{(N+1)^{2}(N+2)^{2}}{H(N+1)} < \frac{(2N-1)H(N)}{N^{2}}.
\end{equation}
Using this relation, we can easily establish that for $N \ge N_{0}$, the sequence $[\pi(N+1)/\pi(N)]_{n}$, $n = 1, 2, \ldots$, is monotonically increasing, is bounded from above,
\begin{equation}
0 < \left[ \frac{\pi(N+1)}{\pi(N)} \right]_{n} < \frac{(N+1)^{2}(N+2)^{2}}
{N^{2}H(N+1)} ,
\end{equation}
and therefore converges. Based on relation (64), we next establish the convergence of the sequence $[\pi(N+1)/\pi(N)]_{n}$, $n = 1, 2, \ldots$, in turn for $N = N_{0}-1, N_{0}-2, \ldots, 1$.

Passing first to the limit $n \rightarrow \infty$  in inequalities (66) and then to the limit $N \rightarrow \infty$, we obtain
\begin{equation}
\lim_{N \rightarrow \infty}\frac{\pi(N+1)}{\pi(N)} = 0, \qquad
\lim_{N \rightarrow \infty} \pi(N) = 0.
\end{equation}
It hence follows that for any value of the parameter $M_{0}$, the components of the field vector $\chi(N)$ found from formulas (58) and (62) tend to zero as $N$ increases infinitely.

If the right-hand side of (62) with $N = 1$ is equated to $H(1)$, then the obtained equation at some fixed value of the parameters $z_{i}$, $\lambda_{i}$, and $c_{0}$ can be regarded as an equation for the mass spectrum of states with
the rest spin 1/2 and parity $r$. When the above equation is satisfied, the quantities $\pi(N)$ obtained using formula (62) satisfy relation (59) for all the required values of $N$, and the corresponding field vectors $\psi(p_{0})$
in (40) are normalizable. For a small deviation of the values of $M_{0}$ from some mass spectrum point $M$, the approximate quantities $\pi(N)_{|_{M_{0}}}$ calculated from (62) and the exact $\pi(N)_{|_{M}}$ satisfying both relation (62)
and equality (60) have similar asymptotic behaviors as $N$ increases and little differ from each other. The degree of smallness of each can then be estimated in appropriate numerical computations.

\begin{center}
{\large \bf 7. Some aspects of analytic calculations of the contributions \\ to
the form factors of lower-rank tensor operators}
\end{center}

The final analytic expressions for electromagnetic form factors (8) and (9), even in the case of calculating only the contributions of lower-rank tensor operators (up to rank four) are quite cumbersome. We therefore restrict ourself to discussing a number of crucial points that lead directly to these 
expressions.

We first note that the spatial parity $r$ in relations (40), (41), (55), and (58) for the state under consideration can be set equal to unity without loss of generality.

The bilinear form $(\psi_{1}, \psi_{2})$ in the space of an $L^{\uparrow}_{+}$ representation decomposable into a direct sum of finite-dimensional irreducible representations is relativistically invariant if [7], [8]
\begin{equation}
(\xi_{\tau' l' m'}, \; \xi_{\tau l m}) = (-1)^{[l]}\delta_{\tau'\tau^{*}} 
\delta_{l'l} \delta_{m'm},
\end{equation}
where $\tau^{*} = (l_{0}, -l_{1}) \sim (-l_{0}, l_{1})$ for 
$\tau = (l_{0}, l_{1})$.

To ensure the realness of form factors (8), we assume that the constants $b_{0}$ and $h_{i}$, $i=0,1,2,3,4$, in formulas (28)--(35) are real. Then
\begin{equation}
(\psi_{1}, \Gamma^{\mu}\psi_{2}) = (\Gamma^{\mu}\psi_{1}, \psi_{2}), 
\qquad (\psi_{1}, L^{\mu}\psi_{2}) = (L^{\mu}\psi_{1}, \psi_{2}), 
\qquad (\psi_{1}, \Gamma^{\mu\nu}\psi_{2})= 
-(\Gamma^{\mu\nu}\psi_{1}, \psi_{2}).
\end{equation}
We hence obtain the equalities
$$(\psi_{1}, \; \Gamma^{\rho}\Gamma^{\mu\nu}\psi_{2}) = 
(\Gamma^{\rho}\psi_{1}, \; \Gamma^{\mu\nu}\psi_{2}), \qquad
(\psi_{1}, \; L^{\rho}\tilde{\Gamma}^{\mu\nu}\psi_{2}) = 
(L^{\rho}\psi_{1}, \; \tilde{\Gamma}^{\mu\nu}\psi_{2}),$$
\begin{equation}
(\psi_{1}, \; \Gamma^{\rho}\Gamma^{\sigma}\Gamma^{\mu\nu}\psi_{2}) = 
(\Gamma^{\sigma}\Gamma^{\rho}\psi_{1}, \; \Gamma^{\mu\nu}\psi_{2}), \qquad
(\psi_{1}, \; L^{\rho}\Gamma^{\sigma}\tilde{\Gamma}^{\mu\nu}\psi_{2}) = 
(\Gamma^{\sigma}L^{\rho}\psi_{1}, \; \tilde{\Gamma}^{\mu\nu}\psi_{2}),
\end{equation}
which considerably simplify both the analytic and the numerical calculations.

We revealed several relations satisfied by the separate terms of form factors (8) and (9), namely,
\begin{eqnarray}
& &\left( \psi_{\frac{1}{2}}(p), \; \Gamma_{i}^{\mu\nu\rho}q_{\nu}q_{\rho}
\psi_{m}(p_{0})\right) = 0, \\
& &\left( \psi_{\frac{1}{2}}(p), \; \Gamma_{i}^{\mu\nu\rho\sigma}q_{\nu}p_{\rho}
q_{\sigma}\psi_{m}(p_{0})\right)=-\frac{1}{2} \left( \psi_{\frac{1}{2}}(p), \; 
\Gamma_{i}^{\mu\nu\rho\sigma}q_{\nu}q_{\rho}q_{\sigma}\psi_{m}(p_{0})\right),
\end{eqnarray}
where $\mu=0,1$, and $i=1,2$. 

The proof of formulas (71) and (72) is based on the transformation properties of the four-vector matrix operators $K^{\mu}$  [7]--[9] and the antisymmetric tensor operators $K^{\mu\nu}$ in (14) under a Lorentz boost along the third 
axis,
\begin{equation}
K^{\mu}p_{\mu}S(\alpha)=S(\alpha)K^{\mu}p_{0\mu}, \hspace{0.3cm} 
K^{03}S(\alpha)=S(\alpha)K^{03}, \hspace{0.3cm} 
K^{1\mu}p_{\mu}S(\alpha)=S(\alpha)K^{1\mu}p_{0\mu},
\end{equation}
and also involves Eqs. (69), the given $P$-parity of the particle state in the rest frame, the relativistic invariance of the bilinear form, and its realness. We demonstrate this in the simplest case where we consider formula (71) with 
$i=1$ and $\mu=0$. We then have
\begin{eqnarray}
&&\left( \psi_{\frac{1}{2}}(p), \; \Gamma^{\rho}q_{\rho}\Gamma^{03}q_{\nu}
\psi_{\frac{1}{2}}(p_{0})\right) = \left( (\Gamma^{\rho}p_{\rho}-
\Gamma^{\rho}p_{0\rho}) S(\alpha)\psi_{\frac{1}{2}}(p_{0}), \; \Gamma^{03}
\psi_{\frac{1}{2}}(p_{0})\right)= \nonumber \\
&& = \left( (S(\alpha)\Gamma^{\rho}p_{0\rho}-
\Gamma^{\rho}S(\alpha)p_{0\rho})P\psi_{\frac{1}{2}}(p_{0}), \; \Gamma^{03}
P\psi_{\frac{1}{2}}(p_{0})\right)= \nonumber \\
&& =\left( (S(-\alpha)\Gamma^{\rho}p_{0\rho}-
\Gamma^{\rho}S(-\alpha)p_{0\rho})\psi_{\frac{1}{2}}(p_{0}), \; P^{2}
(-\Gamma^{03})\psi_{\frac{1}{2}}(p_{0})\right) = \nonumber \\
&& =-\left( S^{-1}(\alpha)(\Gamma^{\rho}p_{0\rho}-
\Gamma^{\rho}p_{\rho})\psi_{\frac{1}{2}}(p_{0}), \; 
\Gamma^{03}\psi_{\frac{1}{2}}(p_{0})\right) = \left( \Gamma^{\rho}q_{\rho}
\psi_{\frac{1}{2}}(p_{0}), \; \Gamma^{03}S(\alpha)
\psi_{\frac{1}{2}}(p_{0})\right)= \nonumber \\
&& =\left( \Gamma^{03}\psi_{\frac{1}{2}}(p), \; \Gamma^{\rho}q_{\rho}
\psi_{\frac{1}{2}}(p_{0})\right) = - \left( \psi_{\frac{1}{2}}(p), \; 
\Gamma^{\rho}q_{\rho}\Gamma^{03}q_{\nu} \psi_{\frac{1}{2}}(p_{0})\right).
\end{eqnarray}

Taking relations (33)--(35) into account, we can show that
\begin{equation}
\Gamma_{2}^{\mu\nu\rho\sigma}q_{\nu}q_{\rho}q_{\sigma}
=c\Gamma_{1}^{\mu\nu\rho\sigma}q_{\nu}q_{\rho}q_{\sigma},
\end{equation}
where $c$ is some numerical coefficient. In the subsequent calculations, we therefore restrict ourself to tensor (6) of the form
\begin{equation}
\Lambda^{\mu\nu}(p,p_{0}) = \Gamma^{\mu\nu} + \Gamma^{\mu\nu\rho}_{1} 
p_{0\rho}+\Gamma^{\mu\nu\rho}_{2}p_{0\rho} + 
\Gamma^{\mu\nu\rho\sigma}_{1}(a_{1}q_{\rho}q_{\sigma}+
a_{2} p_{0\rho}p_{0\sigma})+ \Gamma^{\mu\nu\rho\sigma}_{2}p_{0\rho}p_{0\sigma},
\end{equation}
where $a_{1}$ and $a_{2}$ are arbitrary constants.

We introduce the notation
\begin{equation}
\xi^{\pm}_{l_{1},lm} =
\xi_{(-\frac{1}{2},l_{1})lm} \pm \xi_{(\frac{1}{2},l_{1})lm}
\end{equation}
and also
\begin{eqnarray}
e_{1}(N) &=& \frac{\sqrt{N(N+2)}}{N+1} \chi(N+1) + \frac{2N+1}{N(N+1)} \chi(N) - \nonumber \\
&-& \frac{\sqrt{(N-1)(N+1)}}{N} \chi(N-1), 
\nonumber \\
e_{2}(N) &=& \frac{\sqrt{(N-1)N}}{N+1} \chi(N+1) - 
\frac{\sqrt{(N-1)(N+2)}}{N(N+1)} \chi(N) - \nonumber \\
&-& \frac{\sqrt{(N+1)(N+2)}}{N} \chi(N-1),  
\nonumber \\
e_{3}(N) &=& \frac{\sqrt{N(N+2)}}{N+1} \chi(N+1) - 
\frac{N^{2}+N+1}{N(N+1)} \chi(N) + \nonumber \\
&+& \frac{\sqrt{(N-1)(N+1)}}{N} \chi(N-1), 
\nonumber \\
e_{4}(N) &=& \frac{\sqrt{(N-1)N}}{N+1} \chi(N+1) +
\frac{(2N+1)\sqrt{(N-1)(N+2)}}{N(N+1)} \chi(N) + \nonumber \\
&+& \frac{\sqrt{(N+1)(N+2)}}{N} \chi(N-1) .
\end{eqnarray}

We obtain formulas that essentially simplify the procedure for calculating form factors (8) and (9):
\begin{eqnarray}
& &\Gamma^{1}\psi_{-\frac{1}{2}}(p_{0})=
\frac{c_{0}\sqrt{2}}{6}\left( \sum_{N=1}^{+\infty} \sqrt{2}e_{1}(N)
\xi^{-}_{\frac{1}{2}+N,\frac{1}{2}\frac{1}{2}}+
\sum_{N=2}^{+\infty}ie_{2}(N)\xi^{+}_{\frac{1}{2}+N,\frac{3}{2}\frac{1}{2}}\right),
\nonumber \\
& &\Gamma^{03}\psi_{\frac{1}{2}}(p_{0})=
\frac{h_{0}\sqrt{2}}{3}\left( \sum_{N=1}^{+\infty} \sqrt{2}e_{1}(N)
\xi^{-}_{\frac{1}{2}+N,\frac{1}{2}\frac{1}{2}}+
\sum_{N=2}^{+\infty}ie_{2}(N)\xi^{+}_{\frac{1}{2}+N,\frac{3}{2}\frac{1}{2}}\right),
\nonumber \\
& &\Gamma^{10}\psi_{-\frac{1}{2}}(p_{0})=
-\frac{h_{0}\sqrt{2}}{6}\left( \sum_{N=1}^{+\infty} 2\sqrt{2}e_{1}(N)
\xi^{-}_{\frac{1}{2}+N,\frac{1}{2}\frac{1}{2}}-
\sum_{N=2}^{+\infty}ie_{2}(N)\xi^{+}_{\frac{1}{2}+N,\frac{3}{2}\frac{1}{2}}\right),
\nonumber \\
& &\Gamma^{13}\psi_{-\frac{1}{2}}(p_{0})=
\frac{h_{0}\sqrt{2}}{6}\left( \sum_{N=1}^{+\infty} 2\sqrt{2}e_{3}(N)
\xi^{+}_{\frac{1}{2}+N,\frac{1}{2}\frac{1}{2}}-
\sum_{N=2}^{+\infty}ie_{4}(N)\xi^{-}_{\frac{1}{2}+N,\frac{3}{2}\frac{1}{2}}\right).
\end{eqnarray}

Because the time component of a matrix four-vector operator does not change the spin and its projection on the third axis, it follows that in the calculations of the contributions to form factors (8) and (9) of the third- and fourth-order matrix tensors involved in (76), it is convenient to use relations (70) and the equality
\begin{equation}
\Gamma^{\sigma}\Gamma^{\rho}q_{\sigma}q_{\rho}\psi_{\frac{1}{2}}(p)=
M^{2}[S(\alpha)\Gamma^{0}\Gamma^{0}-2\Gamma^{0}S(\alpha)\Gamma^{0}+ 
\Gamma^{0}\Gamma^{0}S(\alpha)]\psi_{\frac{1}{2}}(p_{0}).
\end{equation}

\begin{center}
{\large \bf 8. Form factors in the theory of ISFIR-class fields with a local
electromagnetic interaction that decrease as $Q^{2}$ increases}
\end{center}

Any statements regarding the possible dependence of electromagnetic form factors (8) and (9) on the transferred momentum squared $Q^{2}$ in the theory of 
ISFIR-class fields are justified from the standpoint of hadron physics if the problem of the mass spectrum is first solved in some version of the theory and a
satisfactory agreement with the experimental picture is obtained for some values of the free parameters. Such a solution, allowing a satisfactory correspondence with the set of observable nucleon resonances, was presented in [14]. The freedom of choice is restricted in the considered version of the theory of 
ISFIR-class fields as follows. The scalar operator $R$ in (53) is determined by two parameters $z_{1}=2.036$ and $z_{2}=0.14$, and we additionally assume that 
$\lambda_{1}/c_{0}=939 \times 2.4686$ MeV and $\lambda_{2}/\lambda_{1}=-0.6724$. In what follows, we consider the state with the lowest mass and the rest spin 1/2.

We first note that for the above values of the parameters of the theory, the magnetic moment of the non-Dirac particle under consideration resulting from the minimal electromagnetic interaction is equal in terms of the magneton not to unity, as for a Dirac particle, but to 0.2824.

We next note that with the specified values of the constants of the theory, the quantities $\rho (N)$ in (53) are positive for all positive integers $N$. Hence, the part of the charged particle electric form factor in (8) ($Q_{0}=1$) that is generated by the minimal electromagnetic interaction, 
$(\psi_{+1/2}(p), R \psi_{+1/2}(p_{0}))$, increases monotonically up to infinity as $Q^{2}$ increases from zero to $+\infty$. Eliminating this increase by introducing a term due to a nonminimal electromagnetic interaction into form factor (8) would cause no concern if the appropriate values of the free constants of the operator $\Lambda^{\mu\nu}$ in (6) had been produced by some mathematically clear procedure.

In the absence of such a procedure, we restrict ourself to establishing the possibility in principle of fixing the numerical parameters of the operator 
$\Lambda^{\mu\nu}$ such that both form factor (8) and form factor (9) ensure a satisfactory approximation of the experimental data for the proton in the domain of small values of the transferred momentum squared ($Q^{2} \leq 0.5$ 
(GeV/c)$^{2}$). Because both form factors (8) and (9) are determined by the same quantity (the tensor $\Lambda^{\mu\nu}$ ), it is not guaranteed a priori that for some choice of the free constants, the $Q^{2}$ dependence of both agrees with the experimentally observed dependence. We can see the validity of this statement in the following example. We assume that $p_{k}=q$ in relation (6) for any index $k$. Then only the even-rank tensors 
$\Gamma^{\mu\nu\nu_{1}\ldots \nu_{j}}$ make nonzero contributions to form factors (8) and (9). As a result of numerical computations with the contributions of tensor operators through the sixth rank taken into account, we find that we can ensure a decrease, which is nearly the dipole one, of the magnetic form factor in (9) as $Q^{2}$ increases. But electric form factor (8) then turns out to be slowly increasing with $Q^{2}$, which contradicts the experimental picture.

The results of approximating the experimental data for the electric and magnetic form factors of the proton obtained using relations (8)--(10), (19), (20), 
(27)--(31), (34), (37), (40), (49), (52)--(54), (58), (62), (68), (70), and (75)--(80) are shown in the figures 1 and 2 with the solid line. They correspond to the constant choices
\begin{eqnarray}
& &\frac{MCh_{0}\sqrt{2}}{3} = 6.9483, \qquad c_{0}h_{1} = -1.4565, \qquad
b_{0}h_{2} = -0.6959, \nonumber \\
& &a_{1} = 0,\qquad c_{0}^{2}h_{3}a_{2} = 0.8008, \qquad 
c_{0}b_{0}h_{4} = -0.1761.
\end{eqnarray}
The dotted lines in the figures also show dipole approximations to the form factors,
\begin{equation}
G_{M} = (1+\kappa)\left( 1+\frac{Q^{2}}{0.71} \right)^{-2}, \qquad
G_{E} = \left( 1+\frac{Q^{2}}{0.71} \right)^{-2},
\end{equation}
where $Q^{2}$ is expressed in units of (GeV/c)$^{2}$. Experimental data are 
taken from [18] and [19].

\begin{center}
{\large \bf 9. Concluding remarks}
\end{center}

The obtained results are largely general and can be useful in investigating this or that theory of non-Dirac fermions both with the rest spin 1/2 and with higher rest spins.

Based on the general structure of matrix second-rank antisymmetric operators given in Sec. 3, we might elsewhere analyze the possibility of the spin operator and the magnetic moment operator of a non-Dirac particle with rest spin 1/2 essentially differing from each other. That, together with the question already
raised in [20] regarding the degree of justification for using 
\newpage

\begin{center}
\vspace{-0.8cm} 
\begin{figure}[ht]
\hspace{0.3cm}
\includegraphics[width=7.6cm]{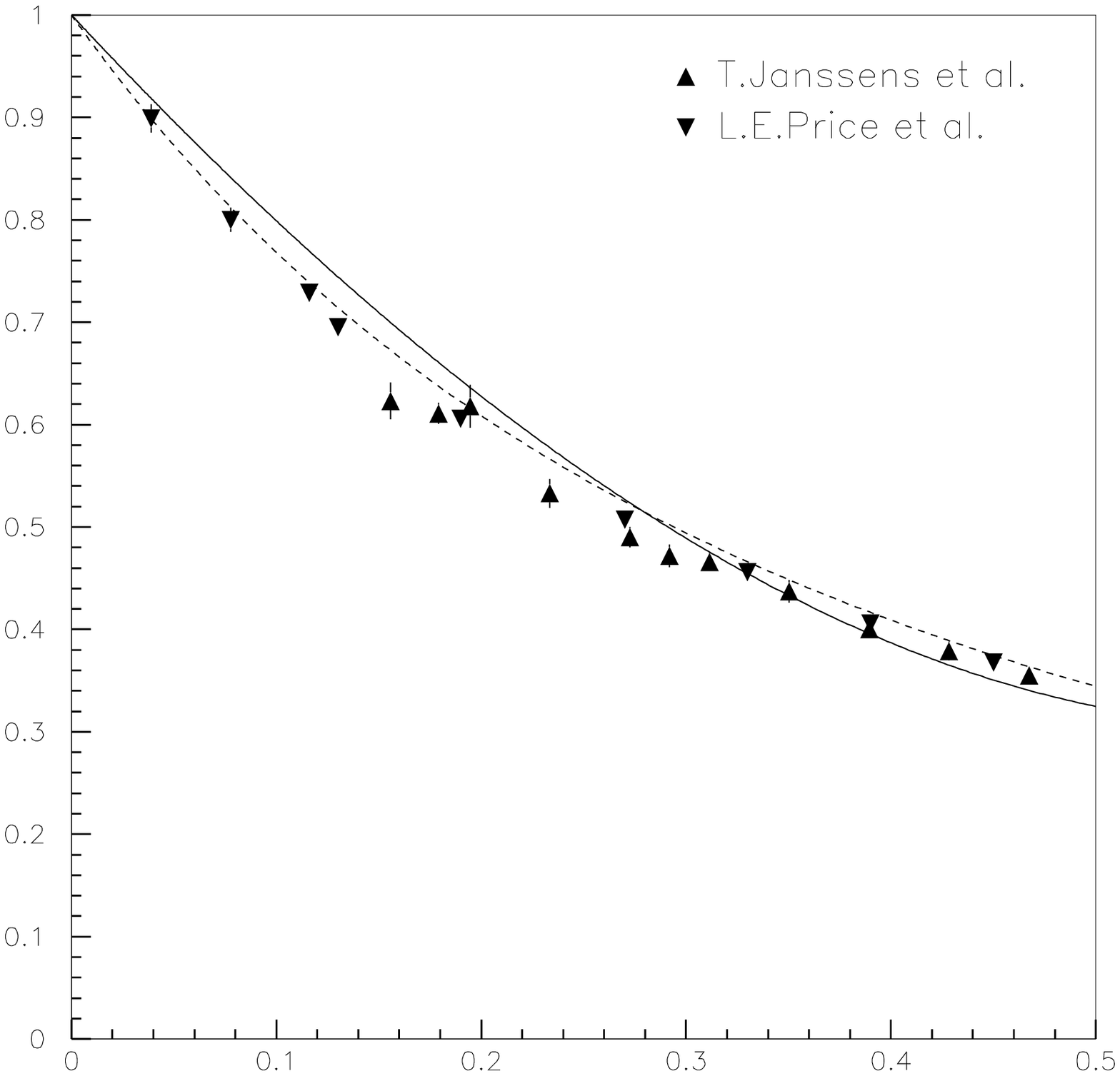}
\includegraphics[width=7.6cm]{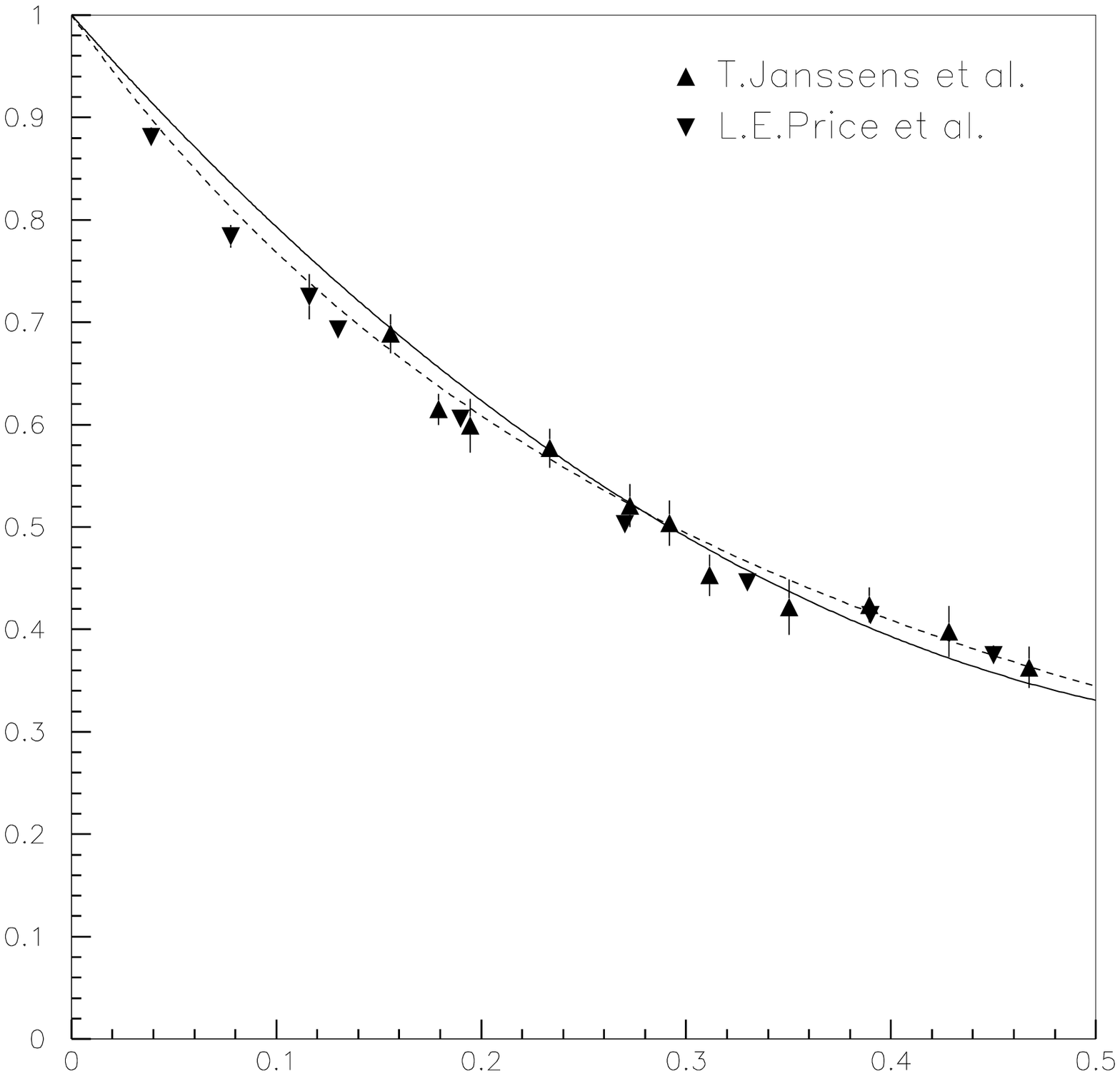}
\vspace*{-1.1cm}
\\
\hspace*{0.55cm}
\parbox[t]{7.0cm}{\caption{The proton magnetic form factor}}
\hspace*{0.5cm}
\parbox[t]{7.0cm}{\caption{The proton electric form factor}}
\end{figure}
\end{center}

\vspace{-9.0cm}
\hspace{-0.7cm} {\small $\frac{G_{M}}{2.79}$} \hspace{7.1cm} {\scriptsize 
$G_{E}$}

\vspace{5.9cm} 
\hspace{5.0cm} {\scriptsize $Q^{2}$ (GeV/c)$^{2}$} \hspace{5.8cm} {\scriptsize 
$Q^{2}$ (GeV/c)$^{2}$}

\vspace{1.0cm}

\noindent
the Bargmann--Michel--Telegdi formula [21] to describe the relativistic particle spin rotation in a constant homogeneous magnetic field, could be an additional argument for the necessity of experimental verification of that formula.

The results of the part of this paper pertaining to the theory of ISFIR-class fields that we investigated do not bury the prospects of luckily assigning the structured particles a local Lagrangian of their electromagnetic interaction. At the same time, the results indicate the existence of nontrivial points requiring
serious thought. We believe it is useful, fully based on the results obtained here and in [9], [13], [14], to analyze the version of the theory of ISFIR-class fields whose free states satisfy the second-order equation
\begin{equation}
\left( c_{1}\partial^{\mu}\partial_{\mu}+ c_{2}\Gamma^{\mu}\Gamma^{\nu}
\partial_{\mu}\partial_{\nu}+i\Gamma^{\mu}\partial_{\mu}-R\right) \Psi(x) = 0
\end{equation}
with known ingredients (the $L^{\uparrow}_{+}$ group representation in (26), the matrix operators $\Gamma^{\mu}$ in (27), and $R$ in (53), (54)) and with some real constants $c_{1}$ and $c_{2}$. This equation can certainly be reduced to a linear equation with spontaneously broken double symmetry, but the newly introduced field is then to correspond to a proper Lorentz group representation whose direct-sum decomposition contains irreducible representations with multiplicities not less than two. Such representations were eliminated from all of our previous investigations by using condition 1 in [9]. Associating Eq. (83) with baryons is supported, in our view, by such a detail in the experimental picture of nucleon resonances [22] as the proximity in masses of the lowest
states with opposite parities for spins 5/2, 7/2, and 9/2. In analyzing the mass spectrum of Eq. (83), it would be desirable to first answer the question of whether a good correspondence with experimental results for nucleons can be obtained in the case where the sequence of matrix elements of $\rho(N)$ in formulas (53) and (54) has an alternating sign (for example, when only one parameter $z_{i}$ is fixed with $z_{1} < -2$). If this possibility is feasible, then we can hope to obtain the minimal interaction contributions to electromagnetic form factors of a charged particle that decrease with an increase in $Q^{2}$ and also to be able to deal with a single matrix second-rank antisymmetric tensor operator for both a charged and a neutral particle.

{\bf Acknowledgments}. The author  is deeply grateful to S.P. Baranov and 
V.E. Troitsky for the numerous discussions of the problems considered in this paper.

\end{small}
\end{document}